\documentclass{aa}  
\usepackage{graphicx}
\usepackage{txfonts}
\usepackage{aalongtable}
\usepackage{subfigure}

\usepackage{natbib}
\bibpunct{(}{)}{;}{a}{}{,}

\input epsf.sty

\begin{document}

   \title{Iron lines in model disk spectra of 
galactic black hole binaries}

   \author{A. R\'o\.za\'nska \inst{1},
          J. Madej \inst{2},
          P. Konorski \inst{3}
          \and 
           A. S\c adowski \inst{1}
	  }
   \offprints{A. R\'o\.za\'nska}

   \institute{  N. Copernicus Astronomical Center,
		Bartycka 18, 00-716 Warsaw, Poland \\
		\email{agata@camk.edu.pl}
                \and
                Warsaw University Observatory, 
                Al. Ujazdowskie 4,
                00-478 Warsaw, Poland \\
                \email{jm@astrouw.edu.pl} 
               \and
             Toru\'n Centre for Astronomy, N. Copernicus University,
             Gagarina 11, 87-100, Torun, Poland \\
             }

   \date{Received ????, 2009; accepted ???, 2009}


\authorrunning{R\'o\.za\'nska et al.}
\titlerunning{Iron lines in model disk spectra of GBHBs}
 
  \abstract
   {We present angle-dependent, broad-band intensity spectra from 
accretion disks around black holes of 10 M$_\odot$. In our computations disks 
are assumed to be slim, which means that the radial advection is taken into account 
while computing effective temperature of the disk.}
   {We attempt to reconstruct continuum and line spectra of X-ray binaries in 
soft state, i.e. dominated by the disk component of multitemperature shape. 
 We follow how the iron line complex depends on the external irradiation, 
 an accretion rate and a black hole spin.}
   {Full radiative transfer is solved including effects 
of Compton scattering, free-free and all important bound-free transitions of 10
main elements. Moreover, we include here the fundamental series of iron lines 
from helium-like and hydrogen-like ions, and fluorescent 
K$_{\alpha}$ and K$_{\beta}$ lines from low ionized iron.}
  {We consider two cases: non-rotating black hole, and black hole  rotating with
almost maximum spin a=0.98, and obtain spectra for five accretion disks from 
hard X-rays to the infrared. In non irradiated disks, 
resonance lines from He-like and H-like iron 
appear mostly in absorption. Such disk spectra exhibit
 limb-darkening in the whole energy range.   
External irradiation causes that iron resonance lines appear
in emission. Furthermore, depending on disk effective temperature, fluorescent
iron K$_{\alpha}$ and K$_{\beta}$ lines are present in disk emitting spectra.
All models with irradiation exhibit limb-brightening 
in their X-ray reflected continua.}
   {We show, that the disk around stellar black hole itself is hot 
enought to produce strong absorption resonance lines of iron. 
Emission lines can be observed only if heating by external X-rays 
dominates over thermal processess in hot disk atmosphere.   
Irradiated disks are usually brighter in X-ray continuum when seen edge on, 
and fainter when seen face on.}
   \keywords{accretion: accretion disks, radiative transfer, X-rays: binaries }

   \maketitle
%

\section{Introduction}

\subsection{General}

The newest X-ray satellite spectra of galactic black hole binaries (GBHBs) 
clearly show the existence of UV/X-ray bump peaking at about 1 keV in their
high/soft states \citep{davis2006}. 
It is widely believed, that this emission originates in an accretion disk circulating 
around the central black hole of stellar-like mass \citep{mitsuda84,dotani97}.
Since individual rings of the accretion disk exhibit 
various effective temperatures,
the radiation spectrum integrated over radii is responsible for multi-blackbody 
shape of the observed spectrum.
Similar optical/UV bumps were observed in the spectra of 
active galactic nuclei (AGN) \citep{shields78,malkan83}.
However, accretion disks in AGN exhibit relatively low effective temperatures, 
since the mass of the central black hole is of the order of $10^8 M_{\odot}$, and 
the disk temperature is inversely proportional to the mass with the power of 0.25.
Multicolor disk model (MCD) still does not explain some spectral details 
\citep{merloni00,kubota2010} 
but up to now there is no better explanation for the observed continuum.  

In addition, in both cases of GBHBs and AGN, substantial fraction of the bolometric
luminosity is emitted in hard X-ray tail extanding up to several tens 
or even hundreds keV \citep{gierlinski97,nandra94}. 
The shape of many X-ray tails shows the evidence 
of interaction between radiation emitted near the central object and
the nearby accreting gas \citep{pounds90}.
The most prominent evidence of such an interaction is
the detection of fluorescent iron K$_{\alpha}$ line at 6.4 keV
in different types of accreting sources \citep{tanaka95,iwasawa99,
miniutti04,miller04,disalvo2005}. 

Shape of an iron line profile carries important information about the 
physics of the material around the compact object.  
Close to the black hole, the emission line profile is relativistically broadened
and skewed, which together with a circular gas movement
in the disk makes the final profile asymmetric with a strong red wing
\citep{fabian89,reynolds97,fabian02}.
But, there is a one example of broad line profile detected in cataclysmic variable, 
GK Per \citep{titarchuk09}, where relativistic effects cannot work at all, 
suggesting that different mechanizm can be responsible for the line broadening. 
Also in some cases, 
the detected iron line profile has narrow component 
\citep{watanabe2003,reeves2004,yaqoob2007}, indicating that interaction 
between hard X-rays and accreting gas occurs farther away from the black hole. 

In all models used to fit X-ray data, iron line profile is assumed to be 
gaussian with eventual kinematic broadening. In this paper we present 
detailed computations of iron line profiles in the case of GBHBs, taking  into 
account thermal, natural, pressure  and Compton broadening of lines when
radiation passes through the atmosphere. Such calculations were already
done in case of the disk in AGN  \citep[][hereafter RM08]{rozanska2008},  
showing profiles of fluorescent K$_\alpha$, K$_\beta$ lines 
and their Compton shoulders.
But accretion disks in GBHBs are hot enough for creation of 
resonance iron lines due to thermal 
processes in the atmosphere. The main goal of this paper is to show 
how strong will be full iron line complex (6.3-7.1 keV)
from hot illuminated disk atmospheres
in GBHBs.  

\subsection{Global disk models}

Numerous theoretical models of the accretion disk spectra were recently 
developed, still a lot of effort reminds to be done to achieve 
undisputable solution. 
One of the obvious reason  for such situation is that each mechanism 
responsible for the accretion makes disk matter very turbulent \citep{ohsuga2007},
and no code at present  can compute radiative transfer 
through moving gas, no matter if this mechanism is driven by 
viscosity 
\citep{SS73}, or by magnetorotational instability \citep{hawley2001}.

Second serious problem is that depending on geometrical and physical assumptions 
we have several global models of accretion disks. The earliest was the standard  
\citet{SS73}, SS disk, which is geometrically thin, and optically thick, with all 
dissipated energy converted into radiation. Relativistic corrections for such disk 
was derived by \citet{novikov73} (hereafter NT disk).
But in other models some thermal energy or mass
in the disk can be carried into black hole by radial advection. 
Such a process generates two branches
of disk solutions: optically thin ADAFs (advection dominated accretion flows)
\citep{narayan1995}, and optically thick, slim disks developed by 
\citet{abramowicz1988}. 
Models of disk spectra presented recently by \citet{davis2006} and storred in  
{\sc xspec} library as {\sc fits} templates, were computed for NT disk. 

We present, for the first time, sample spectra of optically thick, 
slim accretion disks \citep{abramowicz1988}. We trace the behaviour of 
hot and cold (fluorescent) iron line profiles as the function of 
distance from the black hole, accretion rate, and aspect angle. 

\subsection{Modeling theoretical disk spectra}

Theoretical disk spectrum can be obtained on various levels of sofistication, 
from the simple black body emission to the detalied numerical simulation of 
the radiation field. In both cases locally emitted spectrum
 has to be integrated over the disk surface. 
The first approach is quite simple as soon as we know the radial distribution 
of effective temperatures in the disk \citep{SS73}. 
The second approach depends on the assumptions made in calculations of 
vertical structure and radiative transfer in the disk atmosphere. 
The most important aspects which have to be taken into account 
are: treatment of energy-dependent opacities of 
elements, free-free processes together with
Compton scattering, vertical stratification of temperature, density and 
degrees of gas ionization, illumination by  an external X-ray source and
eventual fluorescent emission,
vertical energy dissipation by accretion, eventual influence of vertical
height dependent gravity, which operates in the disk. 
Close to the black hole relativistic corrections on the vertical disk structure
and emitted spectrum should be taken into 
account. 

In the pioneering work done by \citet{laor1989} radiative transfer was 
solved in the Eddington approximation, neglecting Compton scattering.
On the other hand,
\citet{shimura1993} considered fully ionized hydrogen-helium disks
assuming only bremstrahlung and Compton scattering. \citet{dorrer96}
has taken into account some absorption but only that of pure hydrogen. 
None of those papers have treated external illumination. 

On the other hand, there were studies of reprocessing of the external 
hard X-rays by a cold matter, neglecting or simplyfying the treatment  
of hydrostatic equillibrium or detailed source function 
calculations. Reprocessed radiation, created mostly 
by Compton scattering process,  could  be succesfully reconstructed  
by Monte Carlo simulations \citep{george91,magdziarz95}, with true 
absorption taken into account in a very schematic way. 

Reprocessing of radiation by Compton scattering could be described
by Kompaneets equation \citet{ross93,merloni00}. For many years it was the most 
advanced approach to compute the iron line models in the reflected X-ray spectrum. 
But those authors assume that reflection occurs on constant density slab
with optical depth of the order of 3, while thermalization of hard X-rays 
may occur deeper inside an atmosphere. 
Hydrostatic equillibrium was a serious complication of the problem
\citep[as discussed by][and references therein]{rozanska2002}.
Taking into account the constraint of hydrostatic equillibrium
we can precisely calculate strong density and ionization gradient 
in the atmosphere \citep{madej2004}.
 Even if resulting X-ray spectra from constant density slab do 
not differ significantly in the spectral domain where iron line
appears \citep{ross2007},
we point out here that they do not reproduce properly thermal optical 
UV bump, and the strenght of soft X-ray lines, the latter depending on 
ionization vertical profiles.

Very sofisticated calculations of non-illuminated accretion disk spectra were
done with {\sc tlusty} code developed by  \citet{hubeny90}. 
They assume disk atmosphere to be in hydrostatic and radiative equillibrium, 
and solve the radiative transfer equation with non-LTE equation of state
\citep{hubeny97}, 
heavy element abundances and Compton scattering treated with the Kompaneets
equation \citep{hubeny2001}. 
However, those authors do not take into account effects of the external X-ray 
illumination, so they cannot precisely compute iron line profiles,
Nevertheless,
the level of sofistication of our radiative transfer calculations is   
comparable to that presented by \citet{hubeny2001}. 

\subsection{The aim of our paper}

In this paper, we present precise calculations of the radiative transfer through 
the atmosphere of a slim disk around a stellar mass black hole. 
The global slim disk model \citep{olek2010} was computed for  
several values of the accretion rate and for two values of the 
black hole spin: 0 and 0.98.

We use here the code developed by \citet{madej1991} for hot stellar atmospheres
and then adopted to atmospheres illuminated by external hard X-ray source  
\citep{madej2000b}, and to accretion disk geometry 
\citep{rozanska2001}.
Adopting atmospheric computations for an accretion disk, 
we solve the vertical structure and outgoing spectra for several
neighbouring rings being in hydrostatic and radiative equilibrium.   
The final spectrum presented in specific intensity 
scale for different aspect angles is integrated over radii and
presented at the source frame.  No kinematic special relativity
effects  are included in our model. 

Our radiative transfer equation includes effects of multiple Compton 
scattering of radiation on free electrons in relativistic thermal motion,
and rich set of bound-free and free-free opacities. Ionization populations
are computed assuming LTE equation of state for ideal gas. The method allows
for a large relative photon-electron energy exchange at the time of 
Compton scattering and, therefore, are able to reconstruct Compton 
scattering of photons with
energy approaching or even exceeding the electron rest mass \citep{madej2004}. 

We compute profiles of thermal and fluorescent iron lines 
originating from the accretion disk atmosphere 
irradiated by hard X-rays with power-law spectral distribution.
Taking into account emissivity of fluorescent K$_\alpha$ and K$_\beta$ lines 
of low ionized iron in full NLTE, we are able to estimate the relative 
importance of thermal and nonthermal (fluorescent) iron lines.   

In the next section we present 
the global structure of a slim accretion disk model 
model and principal equations used in our radiative transfer computations.
Results are presented in Sec.~\ref{sec:res}, and main conclusions are given in 
Summary section. 

\section{Method and model assumptions}
\label{sec:meth}

\begin{figure}
\epsfxsize=8.8cm \epsfbox[20 150 570 700]{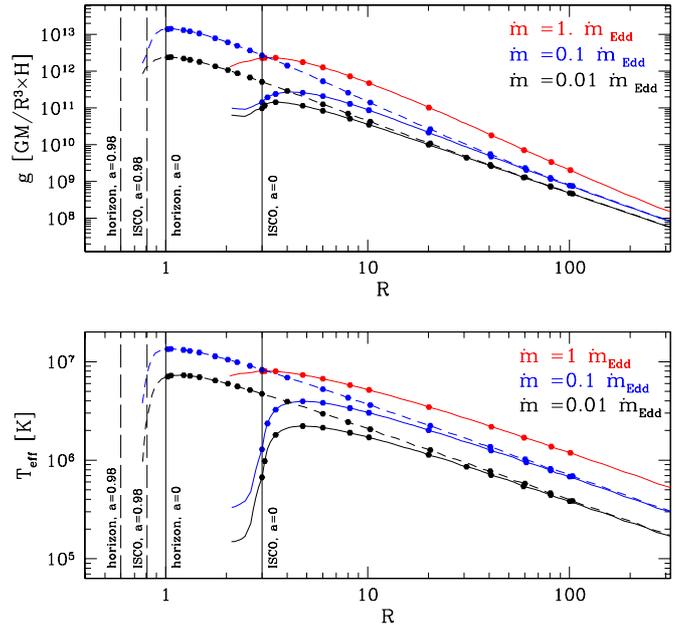}
\caption{Radial dependence of vertical gravity (upper panel) and effective 
temperature (lower panel) in our slim disk models for three accretion rates:
$\dot m=$0.01, 0.1 and 1, denoted as black, blue and red respectiverly.
Two dashed lines (blue and black) represent models for wich we assume 
spin of a black hole equaled 0.98. Thin horizontal lines mark the position
of horizont and ISCO for rotating (dashed line) and non-rotating black hole 
(solid line). Points mark annuli for which detiled radiative transfer computations
were performed.}
\label{fig:global}
\end{figure}

All our models are calculated assuming black hole of the mass $10 M_{\odot}$.
Throught all paper we express radial distances, $R$, in units of the 
Schwarzsild   radius defined as $r_{Schw}=2GM/c^2$.

In the first stage of calculations we solve global two-dimensional, 
hydrodynamical, relativistic slim disk model presented in \citet{olek2010}. 
Contrary to the classical SS and NT disk models it does not assume radiative
equilibrium, i.e. exact balance between the dissipated energy and the 
cooling rate at a given radius. The slim disks allow for advection of heat
with inflowing matter which becomes important for moderate and high accretion rates 
($\dot m > 0.3 \dot M_{Edd}$
\footnote{$\dot M_{\rm Edd} = 16L_{\rm Edd}/c^2$ is the
critical accretion rate that for a disk around a non-rotating BH
approximately corresponds to the Eddington luminosity, $L_{\rm Edd}$.})
and which affects the disk appearence e.g., its effective
temperature and photosphere profiles. The model adopted here \citep{olek2010}
solves the slim disk equations in two dimensions (i.e. assuming only the axial
symmetry). The vertical structure is not assumed ad hoc but is solved consistently.
We assume that the dissipation rate is proportional to the local pressure 
(i.e. we adopt the $\alpha$ prescription with $\alpha=0.01$). 
The energy transport in the vertical 
direction is treated under the diffusive approximation with convection described
by the mixing-length theory. The fraction of the dissipated heat 
which is advected, is independent on the vertical coordinate by assumption.

We computed 3+2 models of accretion disks for the accretion rates 
$\dot m=$0.01, 0.1 and 1 in units of Eddington accretion rate. For all 
values of the acretion rate we consider nonrotating black hole $a=0$,
while for two lowest values of $\dot m$ models  
with $a=0.98$ were also computed.  

Global models considered in this paper are presented in Fig~\ref{fig:global}. 
Upper panel
represents radial distribution of vertical gravity on the disk surface, and lower 
panel - radial distribution of disk effective temperature. 
Both quantities are input parameters in our radiative transfer calculations.
For each global model we divide disk into rings centered at the distances 
shown in Fig~\ref{fig:global}  by rectangle points. For those distances 
detailed radiative calculations were performed.   

As the black hole rotates the innermost stable circular orbit (ISCO) and event
horizont moves closer to the black hole. Therefore, in case of a rotating black hole 
we can receive also radiation from below R=3 (long dashed lines
in Fig~\ref{fig:global}). The latter value defines
the location of ISCO in nonrotating case. 

In all cases, the disk matter is assumed to have  solar-like chemical composition.
Number abundances of elements relative to hydrogen, $(N_{elem}/N_{H})$ equal to:
1.00 (for H), $9.54 \times 10^{-2}$ (He), $4.72 \times 10^{-4}$ (C),
 $9.65 \times 10^{-5}$ (N), 
$8.55 \times 10^{-4}$ (O), $3.84 \times 10^{-5}$ (Ne), $4.17 \times 10^{-5}$ (Mg),  
$4.94 \times 10^{-5}$ (Si), $1.64 \times 10^{-5}$ (S), 
$6.58 \times 10^{-5}$ (Fe). 

\subsection{Radiative transfer calculations}
\label{sec:tran}

Detailed description of the present code and the corresponding equations 
were given in paper RM08, and here we remind only few of them.

The equation of transfer for the specific intensity $I_\nu$ at frequency
$\nu$ is solved in plane-parallel geometry on the monochromatic optical 
 depth $\tau_\nu$:
\begin{equation}
\mu \, {dI_\nu \over {d\tau_\nu}} = I_\nu  - {j_\nu \over {\kappa_\nu +
  \sigma_\nu}} = I_\nu - S_\nu \, ,
\label{eq:dwa}
\end{equation}
where $S_\nu$ is the frequency dependent source function.
In this paper we use the LTE (local thermodynamic
equilibrium) absorption $\kappa_\nu$, whereas coefficients of emission
$j_\nu$ and scattering 
$\sigma_\nu$ include non-LTE terms.
Emission coefficient $j_\nu$ is the sum of three terms, 
$j_\nu = j_\nu^{th} + j_\nu^{sc} + j_\nu^{fl}$, which represent thermal
emission, Compton scattering emission and the emission in iron fluorescent
lines, respectively.

Coefficient of thermal emission $j_\nu$ in LTE is proportional to the
Planck function $j_\nu^{th} = \kappa_\nu B_\nu$ .
 The coefficient of true absorption $\kappa_\nu$, is the sum of
bound-free absorption from numerous levels of atoms 
and ions for all elements, plus free-free absorption from all ions.  
We also included absorption of 4 lowest lines of fundamental series 
of helium-like iron and of similar 4 lowest lines of hydrogen like iron,
all formed in LTE by the assumption.

The external intensity from the point-like lamp is
emitted in the form of power-law with spectral index $\alpha_X$, and 
exponential cut-off limits $\nu_{min}$ and $\nu_{max}$:
\begin{equation}
I_{\nu}^{ext}=A \, \nu^{- \alpha_X} \, \exp\left(- \frac{\nu}{\nu_{max}}\right)
\, \exp\left(- \frac{\nu_{min}}{\nu} \right),
\end{equation}
and  normalized to the luminosity of the source, $L_X$:
\begin{equation}
A  =  \frac{ L_X}{4 \pi r_l^2 \left[  \int_{\nu_{min}}^{\nu_{max}}
\nu^{- \alpha_X} \, \exp\left(- \frac{\nu}{\nu_{max}}\right)  
   \, \exp\left(- \frac{\nu_{min}}{\nu} \right) d\nu \right]} .
\end{equation}
The distance from an X-ray source depends on the ring radius $r$, and the 
lamp height $h_l$,
in the casual relation $r_l^2=h_l^2+r^2$.
The luminosity, spectral index, and cut-off limits of irradiating spectrum
are free parameters of our model, all of them described Sec.~\ref{sec:res}.

Expressions for the Compton scattering terms and emission coefficient in
an irradiated atmosphere were derived in \citet{madej2004} and are valid
in a disk atmosphere without any changes
\vspace{-2mm}
\begin{eqnarray}
j_\nu^{sc} & = & \sigma_\nu J_\nu - \sigma_\nu J_\nu \int_0^\infty 
  \Phi_1 (\nu, \nu^\prime)\, d\nu ^\prime \cr & & + \sigma_\nu \int_0^\infty
  (J_{\nu^\prime}+U_{\nu^\prime})\, \Phi_2(\nu, \nu^\prime)\, d\nu^\prime\, .
\end{eqnarray}

In the above equation variable $U_\nu$ denotes the angle-averaged intensity
of the external irradiation (RM08). 

Compton scattering cross sections were computed following the paper by
\citet{guilbert1981}. Functions $\Phi_1$ and $\Phi_2$ are properly weighted 
angle-averaged Compton redistribution functions for photons both incoming or outgoing
of frequency $\nu$ after scattering in thermal electron gas \citep{madej2004}.

\begin{figure*}
\subfigure{\epsfxsize=0.3\textwidth \epsfbox[10 180 380 700]{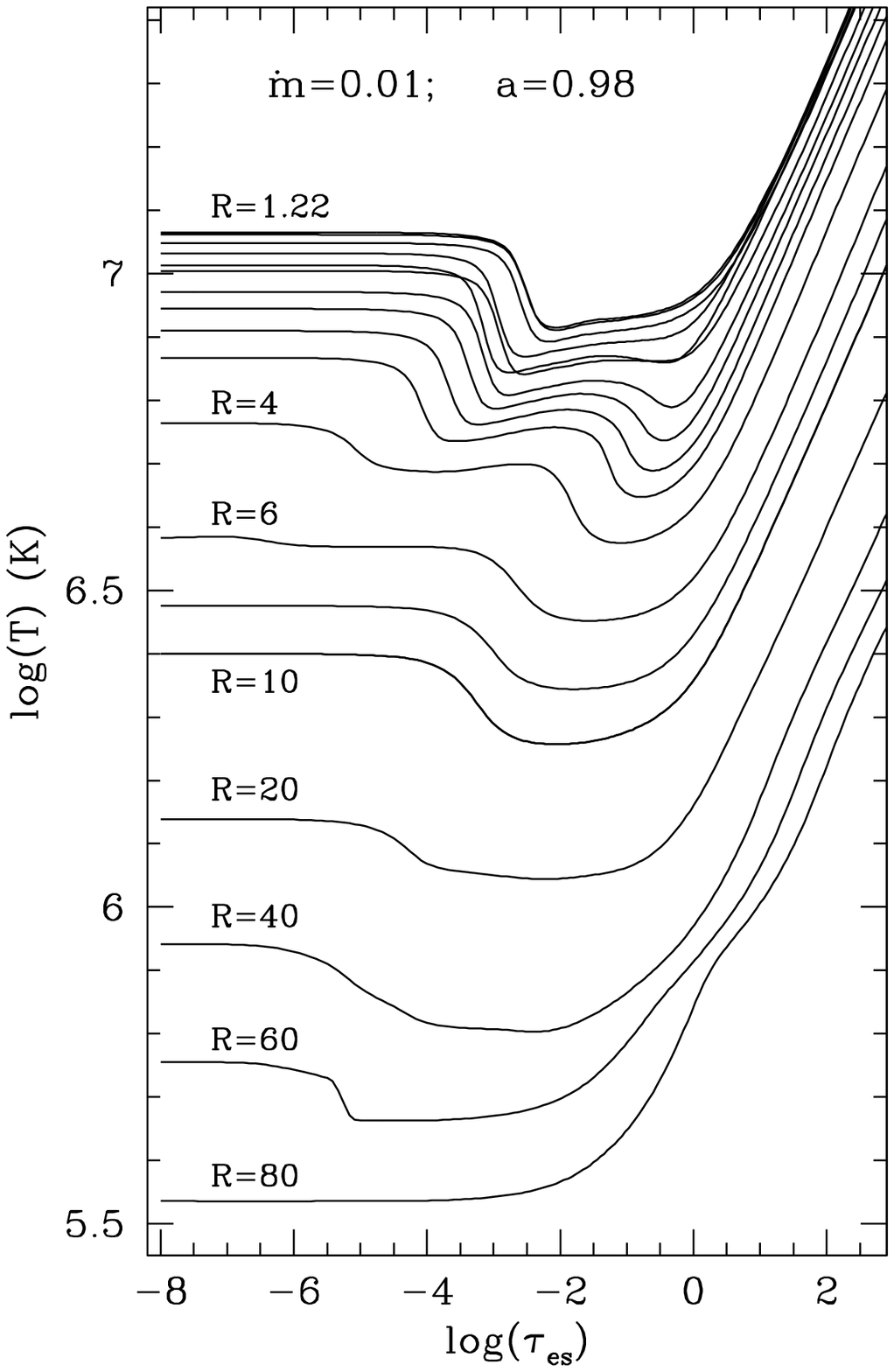}}
\subfigure{\epsfxsize=0.3\textwidth \epsfbox[10 180 380 700]{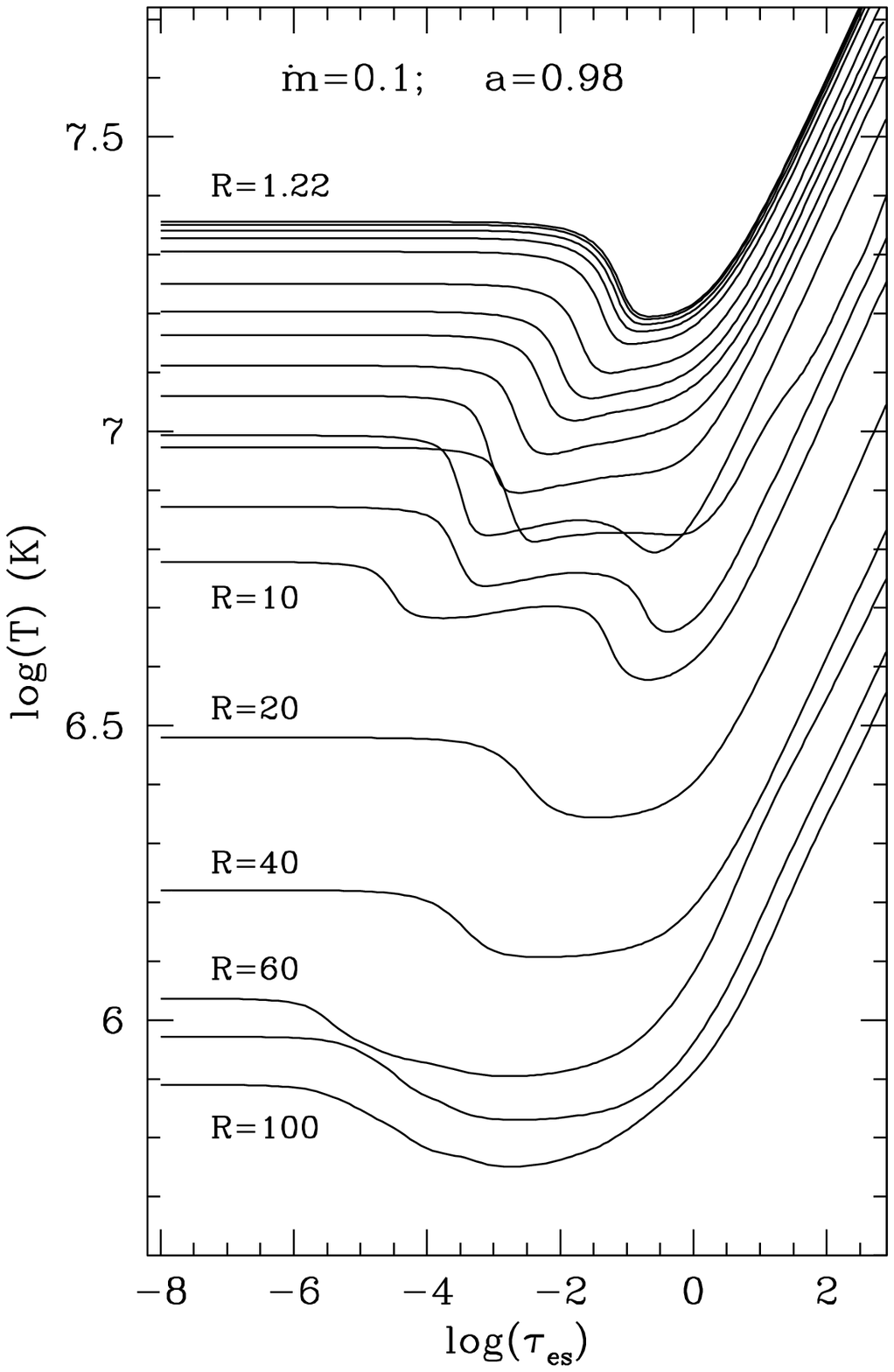}}
\subfigure{\epsfxsize=0.3\textwidth \epsfbox[10 180 380 700]{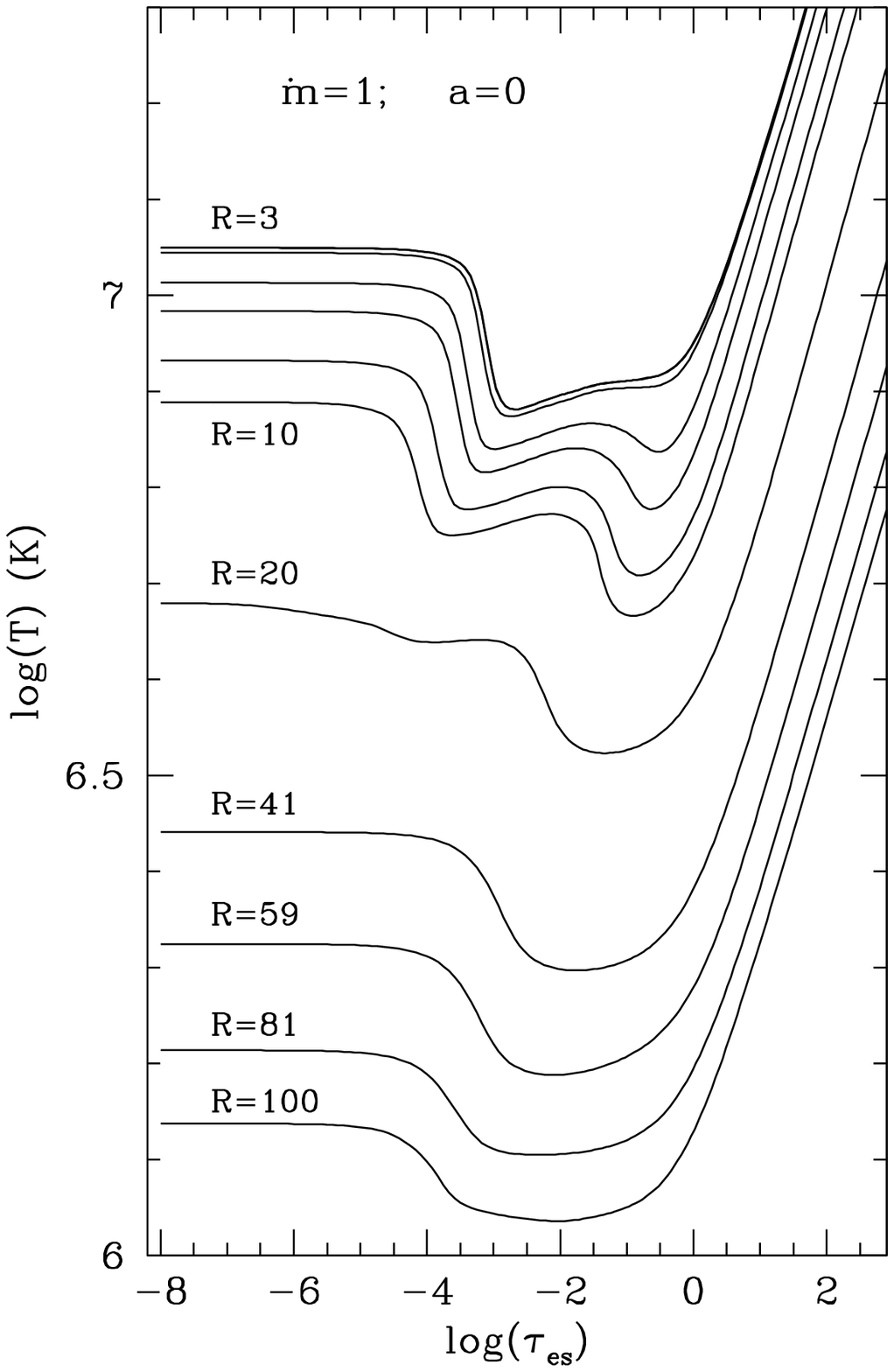}}
\caption{Vertical temperature structure for various rings in three non-illuminated
disk models: Models differ by accretion rate $\dot m$ and dimentionless 
spin of the black hole $a$. Curves are labeled by the distance from the 
black hole, $R$. In all cases we observe temperature rise in outermost 
layers of atmosphere due to Compton scattering (see text for details).} 
\label{fig:fig2}
\end{figure*}

Fluorescence of low-ionized iron gas was approximated by two emission lines,
$K_\alpha$ and $K_\beta$. Therefore,
\begin{equation}
j_\nu^{fl} = E_\alpha^{fl} \, \varphi_\nu^\alpha + 
   E_\beta^{fl} \varphi_\nu^\beta \, ,
\end{equation}
where $E_\alpha^{fl}$ and $E_\beta^{fl}$ denote the integrated intensity
of $K_\alpha$ and $K_\beta$ emission lines, respectively. Frequency
dependent variables $\varphi_\nu^\alpha$ and $\varphi_\nu^\beta$ define
profiles of fluorescent lines, both normalized to unity.

Energy emitted in $K_\alpha$ and $K_\beta$ lines was derived by absorption
of hard continuum X-rays from the radiation field  penetrating the disk
atmosphere. Hard X-ray photon interacting with neutral or low-ionized iron
most probably ionize and remove electron from the innermost $K$ shell, and
then remaining hole is filled by another electron falling from $L$ shell 
($K_\alpha$) or $M$ shell ($K_\beta$ transition). 
The integrated emissivity for Fe $K_\alpha$ fluorescent line is given by:
\begin{equation}
E_\alpha^{fl} = Y \times h\nu_0 \int_{\nu_0}^\infty
  {\alpha_\nu^{\rm iron} \over {h\nu}} (J_\nu + U_\nu) \, d\nu \, ,
\end{equation}
where we set $Y=0.34$ \citep{krause1979}. Variable $\alpha_\nu^{\rm iron}$
is the bound-free absorption coefficient for ionization from $K$ shell
of iron counted for 1 atom. Similar expression holds for the integrated
intensity $E_\beta^{fl}$.

The fluorescence yield, $Y$, defines the fraction
of the energy of hard X-rays absorbed by iron atoms which was reemitted as
photons in $K_\alpha$ line. The remaining energy of absorbed X-rays,
$1-Y$, was spent for the release of Auger electrons.

Iron $K_\alpha$ fluorescent line is a doublet line,
and such a structure was reproduced by our code.
The doublet structure is caused by the fact that L shell has three sub-shells 
depending on the value of the spin and orbital angular momentum, and only two 
transitions are allowed by selection rules. 
We set the central energies for $K_{\alpha_1}$ and
$K_{\alpha_2}$ lines to 6.404 keV and 6.391 keV, respectively. 
Natural widths (FWHM) of both lines, 2.7 eV ($K_{\alpha_1}$) and 3.3 eV 
($K_{\alpha_2}$) were taken from \citet{krause1979}.
The $K_\beta$ line was approximated by a singlet line at the energy
7.057 keV. We set the natural width of the line to 2.5 eV arbitrarily.

Integrated emissivity $E_\alpha^{fl}$ was divided in the proportion 2:1 
between $K_{\alpha_1}$ and $K_{\alpha_2}$ components of the doublet,
since the number of electrons on L shell with total spin equaled 1 is twice
higher than the number of electrons with total spin equal 0. Following 
\citet{basko1978} we quite arbitrarily assumed that the integrated
intensity $E_\beta^{fl} = 0.1 \, E_\alpha^{fl}$.
Opacity profiles of all three lines were set to Voigt functions with 
depth-dependent parameters describing natural and thermal broadening.

On each discrete ring we performed radiative transfer calculations assuming 
that the disk atmosphere remains in hydrostatic and radiative equillibrium
(Eq. 11 and 12 in RM08). 
No turbulences and convection were taken into account.

Full set of model atmosphere equations is solved by the method of 
partial linearization \citep{madej2004}. Equation of hydrostatic equillibrium was 
excluded from the linearization scheme. The equation of transfer, Eq.~\ref{eq:dwa} 
is solved by the Feautrier method and the variable Eddington factors method
\citep{mihalas1978}.  

\begin{figure*}
\subfigure{\epsfxsize=0.3\textwidth \epsfbox[10 180 380 700]{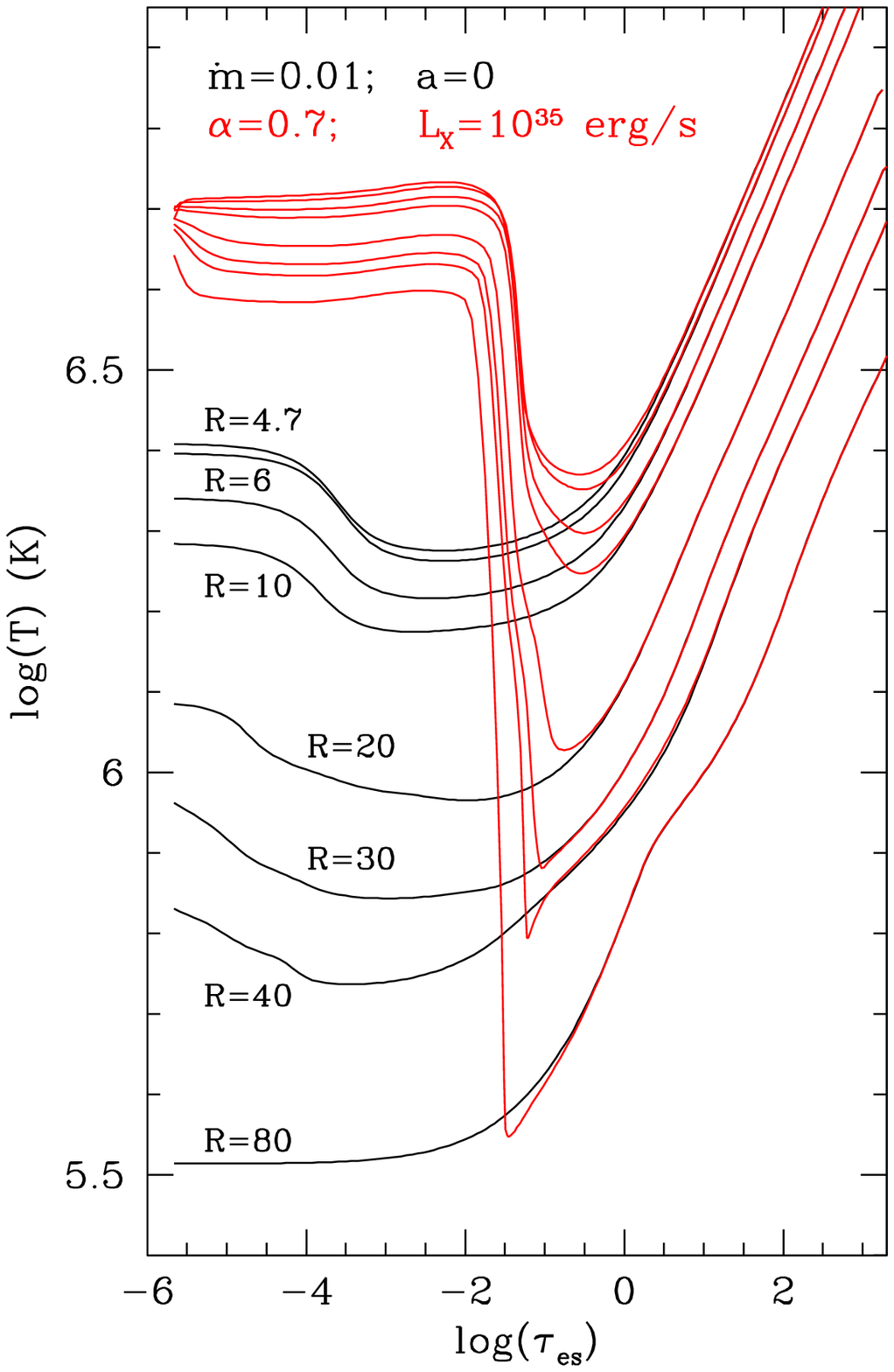}}
\subfigure{\epsfxsize=0.3\textwidth \epsfbox[10 180 380 700]{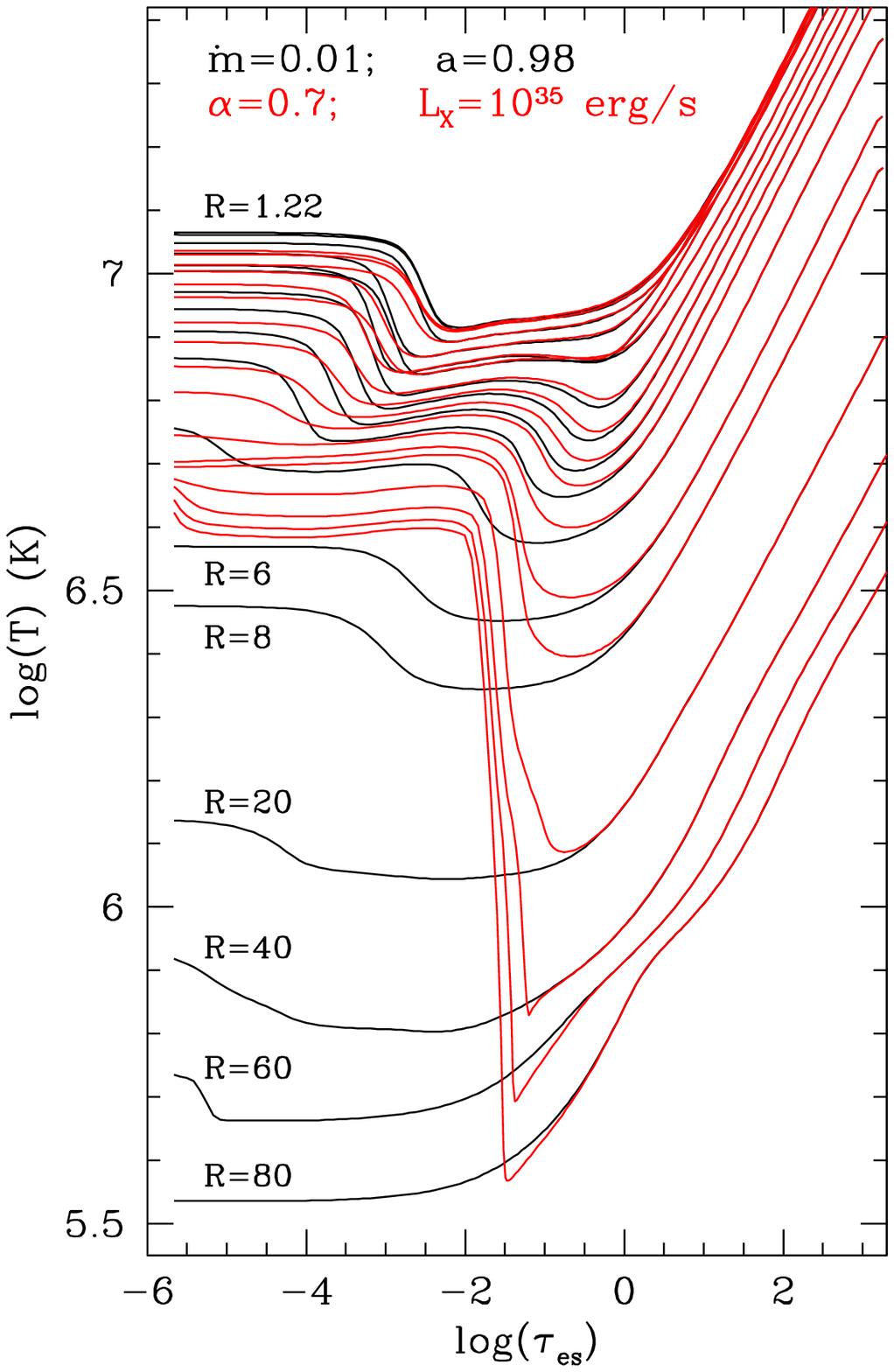}}
\subfigure{\epsfxsize=0.3\textwidth \epsfbox[10 180 380 700]{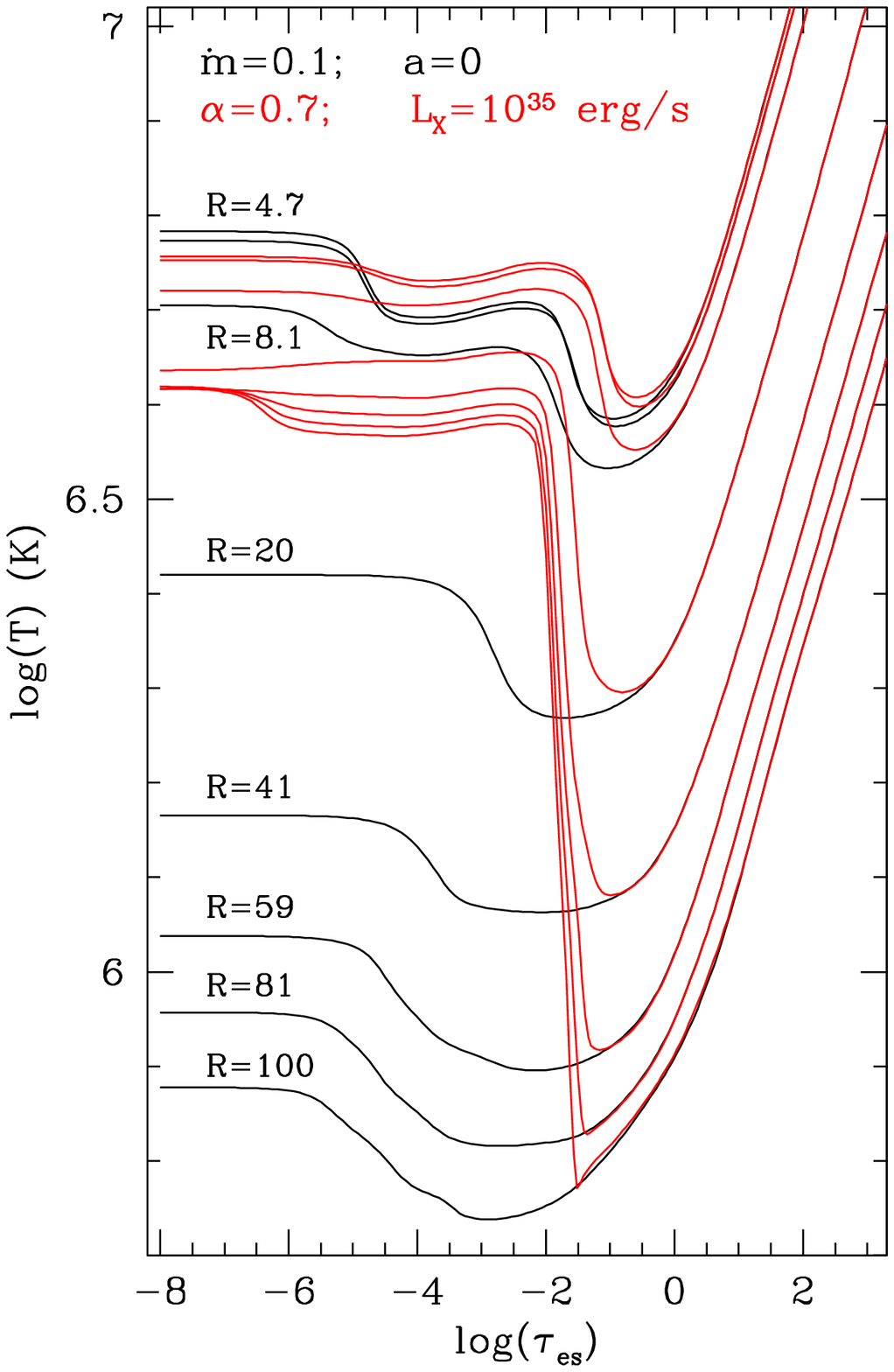}}
\caption{Comparison of the temperature structure in three disk models before (black line)
and after (red line)  illumination by external radiation in the form of power-law.
Results are shown for various distances from the black hole.
It is clearly seen that the external illumination affects only outermost layers of the disk
atmosphere. The most prominent  effects of irradiation occur at distances 
where the effective temperature of an atmosphere itself is relatively low. 
It happens always when the  accretion rate is low, see left panel. 
For high accretion rate, high spin and close to the black hole, the atmosphere 
itself is hot and the external illumination can slighly cool the outermost layers due to 
Compton cooling.}
\label{fig:temosw}
\end{figure*}

\section{Results}
\label{sec:res}

Our code allows us to compute the structure of disk atmospheres
over very large range of electron scattering optical depth starting from
$\tau_{es}= 10^{-8}$ up to $\tau_{es}= 10^{5}$. Furthermore, we were able
to reproduce the overall continuum spectrum from deep infrared of 0.4 eV
up to hard X-rays of 400 keV. 
We solve the radiative transfer problem on the grid of 175 optical depth and 
2200 photon energies simultaniously. 

All spectra are presented as energy dependent 
outgoing specific intensities, $I_E$, which are suitable for disk geometry.
We reject presentation of monochromatic fluxes since they are relevant only
to geometry of a spherical star. Our intensities are shown 
for 8 cosines of viewing angles, which represent angles of the Legendre quadrature. 
Exact values of those angles, their cosines and the type of lines used
in all figures presenting disk spectra 
are given in Table~\ref{tab:cosin}. In further discussion we draw 
attention of the reader to the extreme angles: solid black line represents
almost vertical direction (face-on aspect), whereas the solid red
line represents almost horizontal direction (edge-on aspect).

\begin{table}
\begin{center}
\caption{Description of lines in  Figs.~\ref{fig:fig3} -~\ref{fig:totliosw}}
\label{tab:cosin}
\begin{tabular}{cccccc} 
\hline \hline 
type of a line & $\cos(i)$ & $i$ \\\hline
solid black  & 0.9801 &  $11.4^{\circ} $ \\
short-long dashed &  0.8983 &  $26.1^{\circ} $ \\
long dashed dotted   &  0.7628 &  $40.3^{\circ} $\\
short dashed dotted   &  0.5917 &  $53.7^{\circ} $\\
dotted   &  0.4083 & $ 65.9^{\circ} $ \\
short dashed   &  0.2372 &  $76.3^{\circ} $ \\
long dashed  &  0.1017 &  $84.2^{\circ} $\\
solid red &  0.0199 & $88.9^{\circ} $\\
\hline \hline
\end{tabular}
\end{center}
\end{table}

We consider only one choice of irradiating power-law with  $\alpha_X=0.7$
and luminosity  $L_X=10^{35}$ erg~s$^{-1}$, ranging from  $h \nu_{min} = 0.1 $
up to  $h \nu_{max} = 100$ keV. Position of the X-ray lamp is fixed arbitrarily
on the height $h_l=5 r_{Schw}$ above the disk at the radius  $R=2$
from the black hole. 

\subsection{Temperature structure}
\label{sec:temp}

Radial distribution of effective temperatures in all global models 
has characteristic shape (see Fig.~\ref{fig:global}. 
For non rotating black hole the maximum value of 
the effective temperature occurs on $R \approx 5$. Closer to the black hole, 
there is a slight temperature drop, which is stronger for the lower accretion 
rate. For $\dot m =0.01$, the effective temperature at R=3 has the same 
value as on R about 60, while the surface gravity is sligtly higher. 
This matters in case of irradiated atmospheres, where the fluorescent line 
can be created due to illumination of innermost ring with the effective temperature 
of the order of $7 \times 10^5$ K (see Sec.~\ref{sec:indiv}).
Moving farther from the black hole temperature decreases, and at R=100 
is half order of magnitude lower than its maximum value. 
When the black hole rotates the temperature of innermost rings increases 
reaching $10^7$ K, and we can expect high ionization degree of iron in
such conditions. 

Vertical temperature profiles computed for individual rings are presented
for various global models and distances at Fig.~\ref{fig:fig2} and~\ref{fig:temosw}.
All profiles of the disk atmospheres exhibit inverse temperature rise in 
the uppermost atmospheric layers. 
In the case of non-illuminated disks, presented 
in Fig.~\ref{fig:fig2}, the inverse 
temperature rise is due to Compton heating by hot X-ray photons from below the photosphere.
This result is in agreement with models of hot neutron star atmospheres computed 
by \citet{majczyna05,suleimanov07}, and with atmosphere of accretion disks
presented by \citet{hubeny2001}. Note that in Fig.~\ref{fig:fig2}, 
we present three hottest models where the surface temperature reaches $2.2 \times 10^7$ K. 
In such conditions, iron is almost fully ionized due to heating by 
radiation generated already via accretion process. 

In disks with external illumination the heating effect is enhanced 
(see red lines in Fig.~\ref{fig:temosw})
due to Compton scattering of hot external X-ray photons. 
The amount of this heating depends on the disk effective temperature determined by
the accretion rate and the black hole spin. 
For the lowest accretion rate and non rotating black hole (left panel)
the effect of irradiation is seen in all rings, since effective temperatures for this model
are the lowest. When the black hole rotates
(middle panel) X-rays heat up mostly outer rings. The same is observed for 
higher accretion rate and non rotating black hole (Fig.~\ref{fig:temosw} right panel).
We do not show temperature run for two hottest disks, since then the external irradiation
does not cause any significant heating effect. 

Our results include the influence of both Compton heating sources: 
by hot photons from the photosphere, and from the external X-ray source. 
Because of that, surface temperature structure depends on the shape and 
the normalization of the external irradiation. In case when the atmosphere is hot 
by itself, additional soft X-ray photons can cool down the hottest layers 
due to Compton down-scattering. This happens for rings located close to the 
black hole, for the model with $\dot m=1$ and $a=0$, 
where the temperature of the uppermost irradiated surface layers is lower than for  
non-irradiated atmosphere (Fig.~\ref{fig:temosw} right panel).
Compton down-cattering is mostly caused by photons with energies 
lower than  gas temperature, so we can expect that hot disk atmospheres
in galactic black hole binaries can be cool down due to irradiation by 
soft X-rays. We plan to study this effect in a forthcoming paper. 

\subsection{Spectra from individual annuli}
\label{sec:indiv}

\begin{figure*}
\subfigure{\epsfxsize=0.24\textwidth \epsfbox[30 180 330 700]{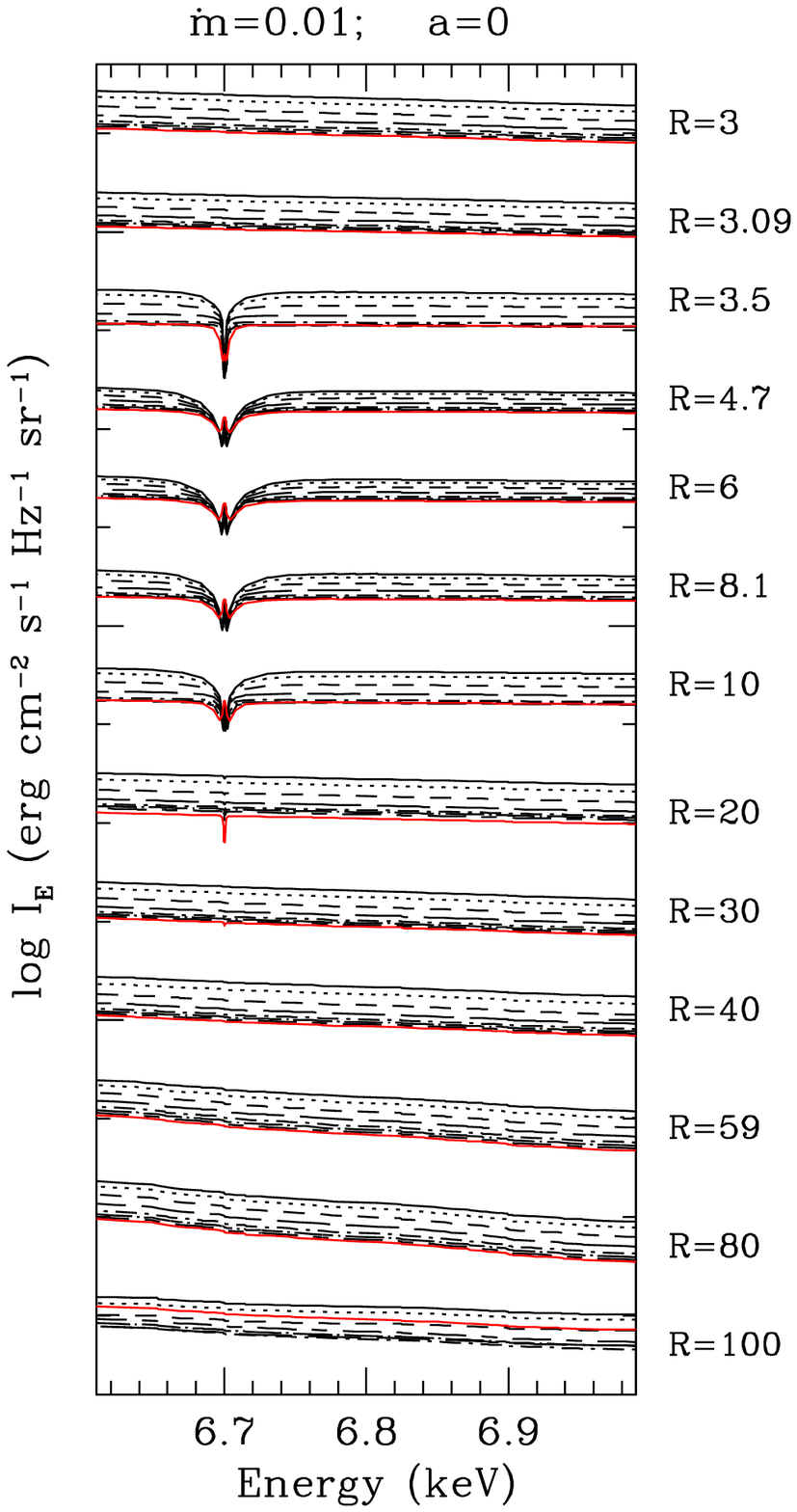}}
\subfigure{\epsfxsize=0.24\textwidth \epsfbox[30 180 330 700]{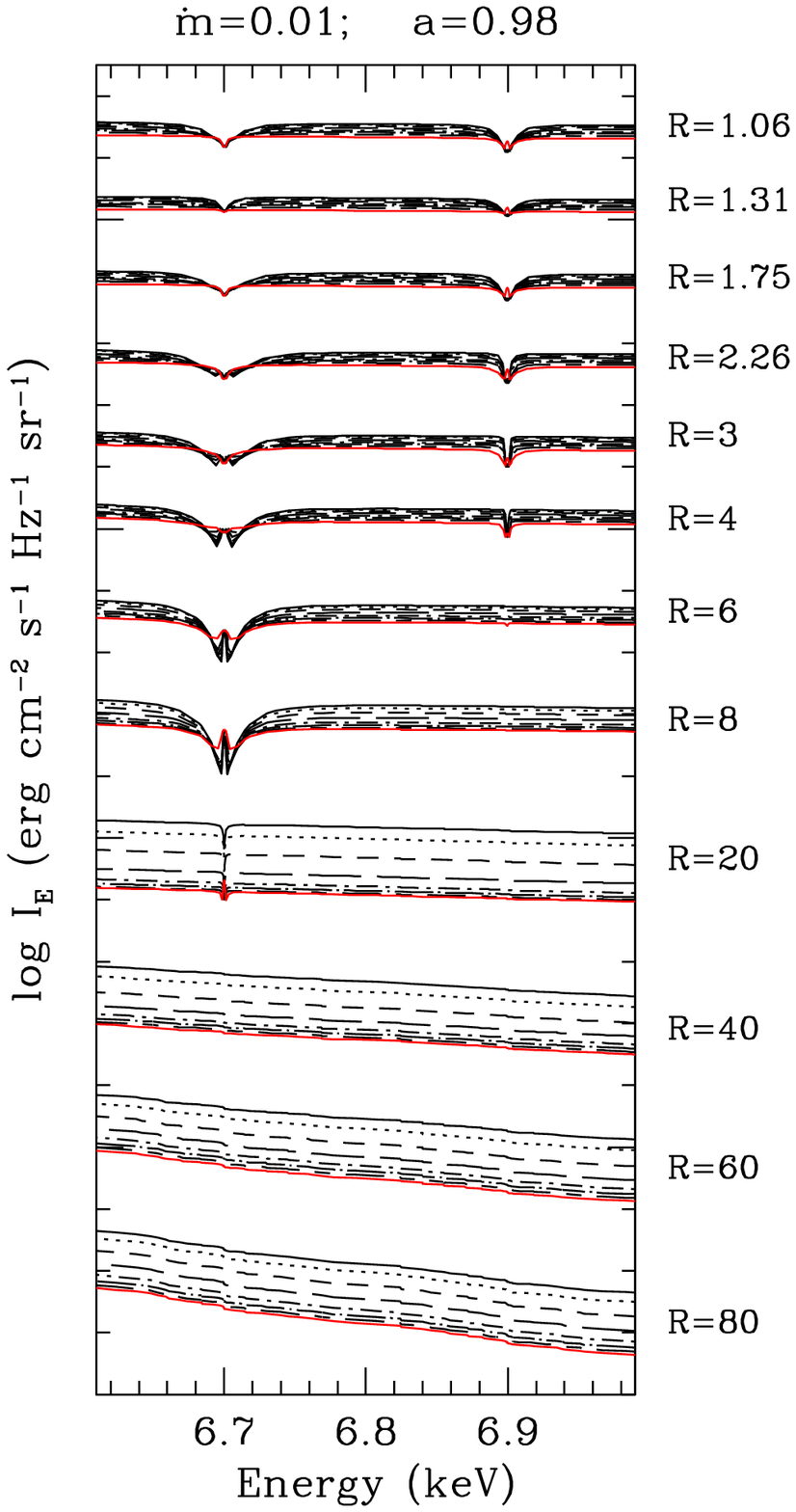}}
\subfigure{\epsfxsize=0.24\textwidth \epsfbox[30 180 330 700]{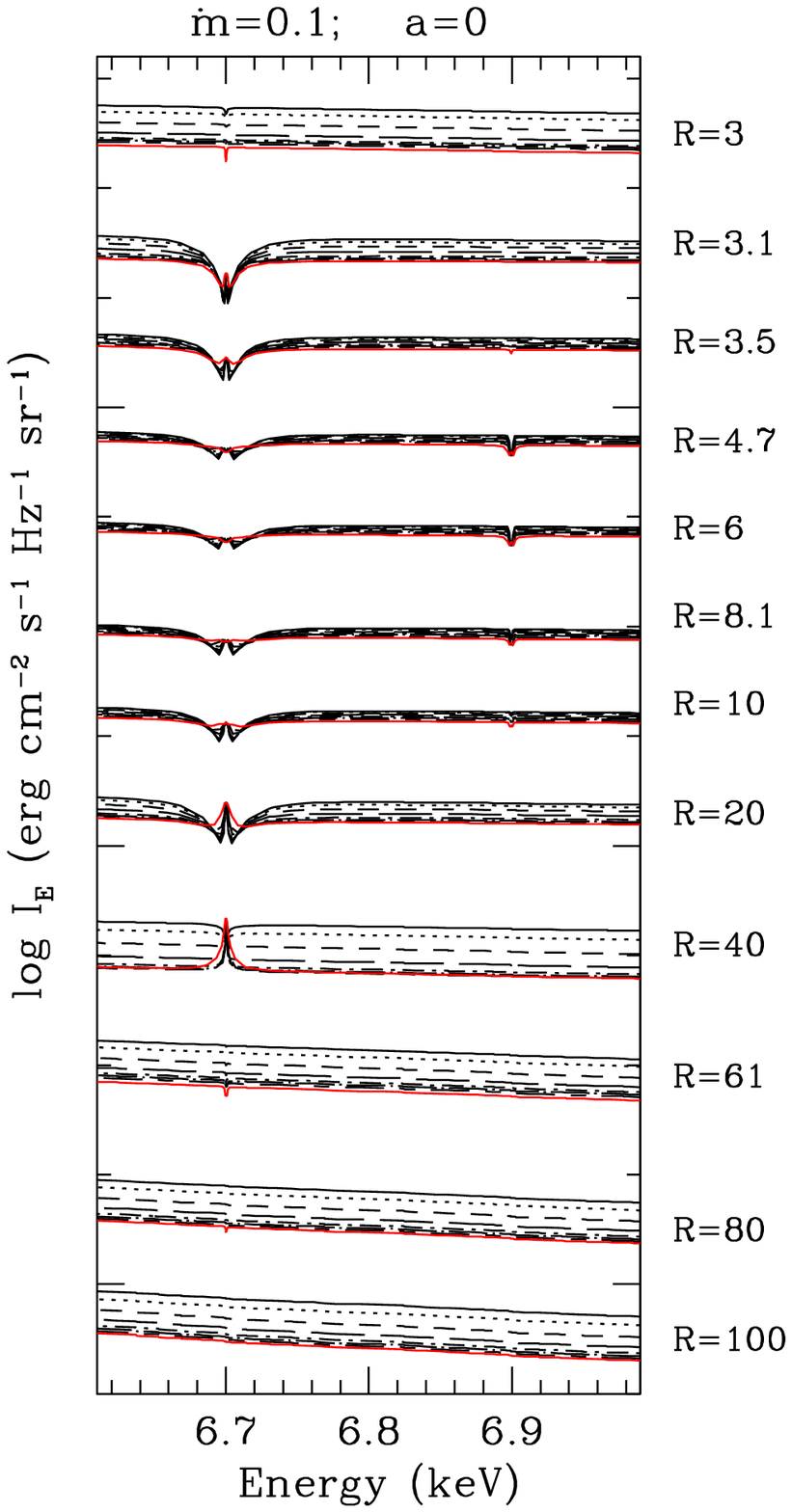}}
\subfigure{\epsfxsize=0.24\textwidth \epsfbox[30 180 330 700]{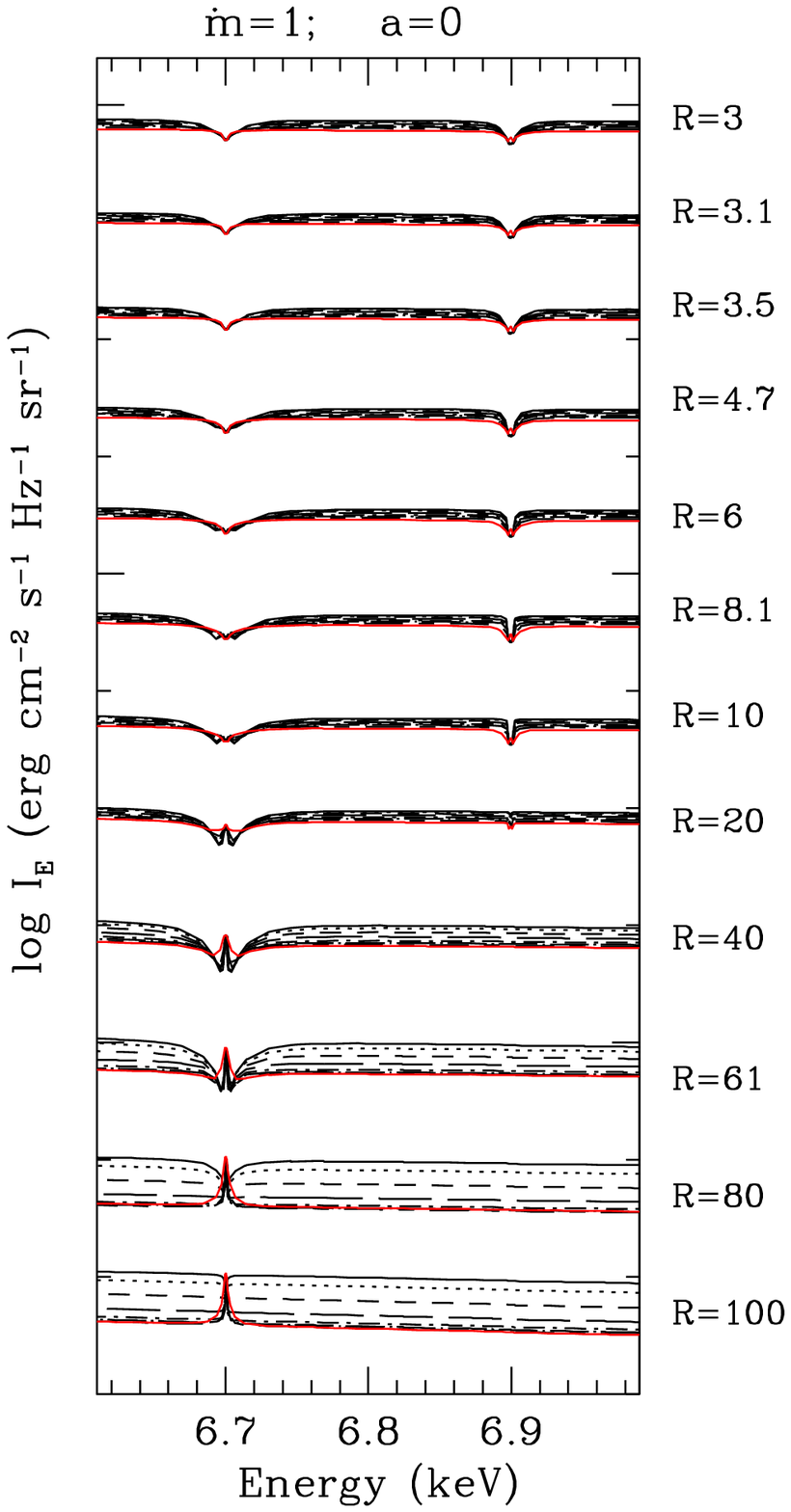}}
\caption{Outgoing intensity spectra from individual rings in four non illuminated 
disk models at different accretion rates and black hole spins. 
Spectra are shown in the energy range corresponding to the resonanse lines 
from hydrogen and helium like iron. Spectra are shifted along vertical axis 
in order to demonstrate evolution of lines with distance  from the black hole 
(marked on the right vertical axis). The limb-darkening effect,
where intensity from ``face-on'' disk (black solid line) is higher
than intensity from ``edge-on'' disk (red solid line), is observed in all 
cases, except in some line cores.}
\label{fig:fig3}
\end{figure*}

Intensity spectra emitted by individual rings of non irradiated accretion disks 
are presented in Fig.~\ref{fig:fig3}. Each panel shows evolution 
of resonance lines from He-like and H-like iron for four models with 
the lowest effective temperatures.  
Line profiles change with distance from the black hole, which is marked on the 
right vertical axis. 
In some rings the line at 6.7 keV exhibits reverse emission in the line core which
(for LTE) is the signature of the temperature inversion.

\begin{figure*}
\subfigure{\epsfxsize=0.31\textwidth \epsfbox[30 170 370 700]{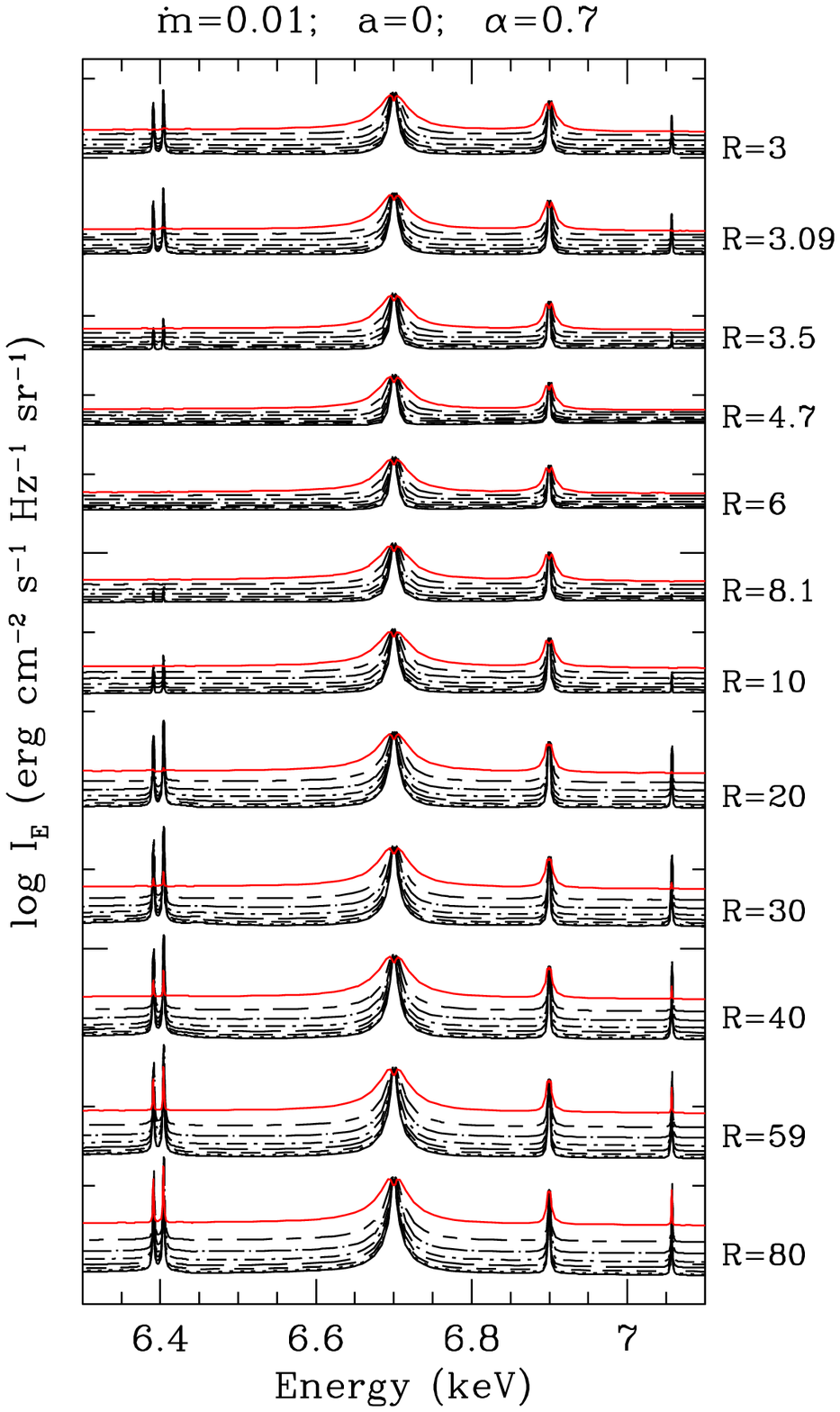}}
\subfigure{\epsfxsize=0.31\textwidth \epsfbox[30 170 370 700]{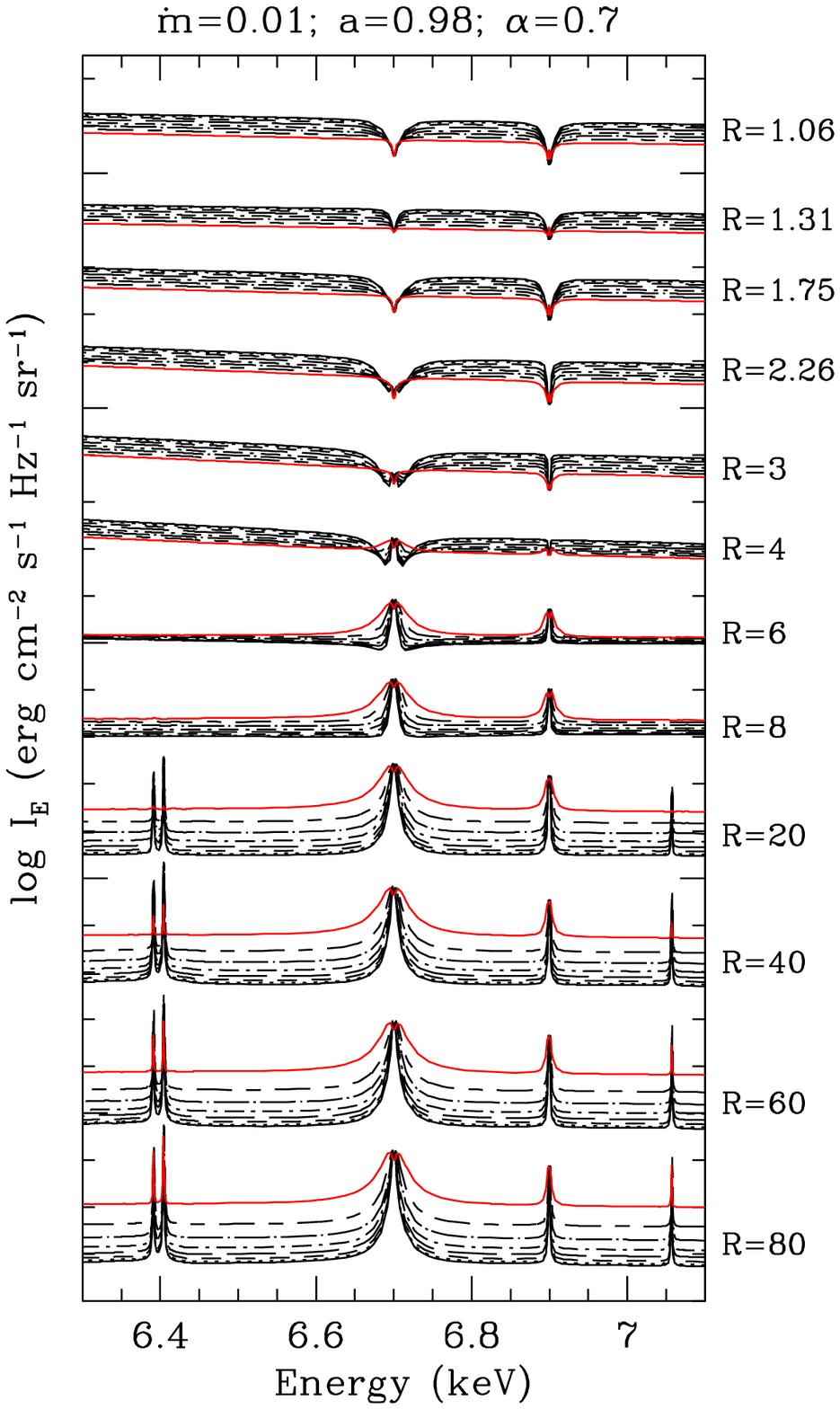}}
\subfigure{\epsfxsize=0.31\textwidth \epsfbox[30 170 370 700]{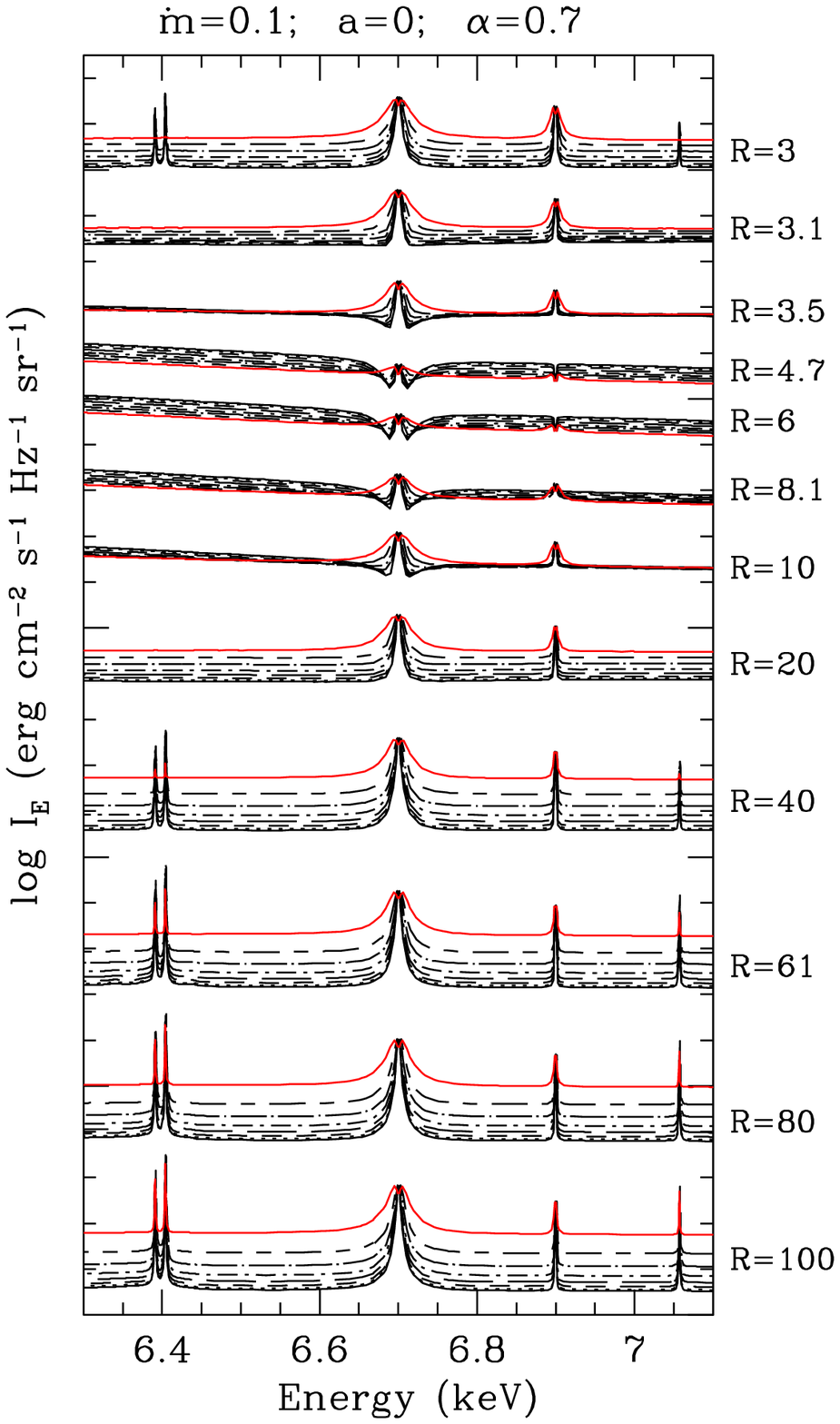}}
\caption{Outgoing intensity X-ray spectra for sample of irradiated accretion disks.
Spectra are shifted along vertical axis in order to demonstrate the evolution of iron 
line complex with distance from the black hole (marked on the right vertical axis).}
\label{fig:liosw}
\end{figure*}

For the disk with the lowest accretion rate and zero spin, only absorption line
from He-like iron is present originating in rings of the maximum effective temperature. 
Spectra from two rings located near the black hole, and
from all rings above $R=20$ are featureless, since temperature is low and 
there is not enought He-like and  iron ions. Situation changes for the same 
accretion rate when the black hole 
rotates. Iron line at 6.7 keV is stronger and the line at 6.9 keV becomes visible. 
For $\dot m=1$ and $a=0$ both lines are clearly seen for most of radii $R$.
There exists energy dependent limb-darkening in all non irradiated models. 
At some rings, differences between intensities emitted at  various
angles are strong, while sometimes there are not seen at all. 

Spectra of irradiated disks are presented in Fig.~\ref{fig:liosw} for the energy
range 6.3-7.1 keV. For most rings both resonance lines of iron 
appear in emission. This effect, caused by external illumination is visible 
only for those rings, when there is temperature change due to irradiation 
(see also Fig.~\ref{fig:temosw}). 
For the model presented in the middle panel for $R<6$, both resonant lines appear
in absorption. 

Rings with relatively lower effective temperature, usually located 
farther from the black hole 
show cold fluorescent iron lines, K$_{\alpha}$ at 6.4 keV and K$_{\beta}$ at 7.05 keV. 
We point out here that those lines can be visible in the whole  
iron line complex, when integrated over disk surface (see the following subsection).

External irradiation causes limb-brightening in intensity spectra emitted from 
numerous rings. 
In the energy domain of Fig.~\ref{fig:liosw} continuum limb-brightening apparently 
is correlated
with the reverse emission in both resonance lines of iron. 
For the model presented at the middle panel,
lines emitted from innermost rings are in absorption and 
simultaneously we observe usual limb-darkening. 

\subsection{Disk-integrated spectra} 

\begin{figure}
\epsfxsize=8.8cm \epsfbox[80 150 580 700]{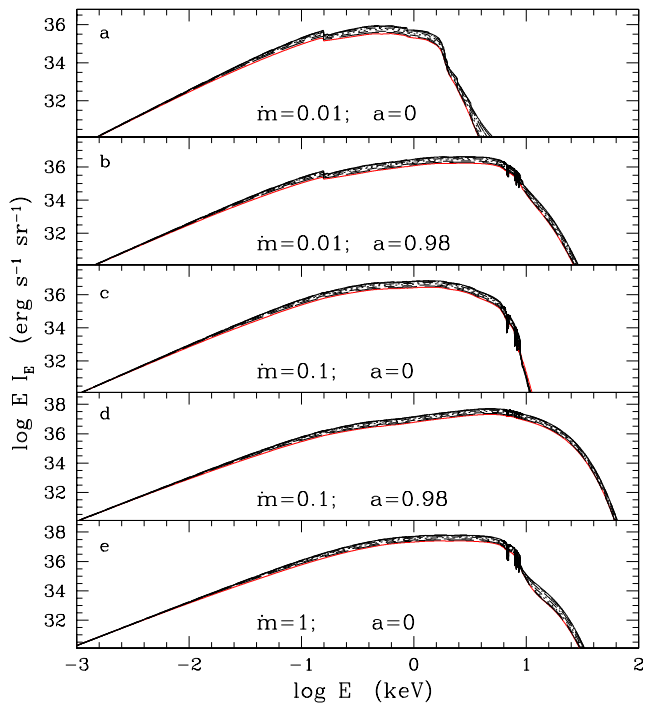}
\caption{Broad-band intensity spectra integrated over disk surface
for all non-illuminated disk models for 
eight viewing angles. The shape of continuum is affected by several spectral edges and 
iron line complex around 6.9 keV. Depending on the accretion rate and spin of the 
black hole we can follow how limb darkening changes with energy.}
\label{fig:total}
\end{figure}

\begin{figure}
\epsfxsize=8.8cm \epsfbox[80 150 580 700]{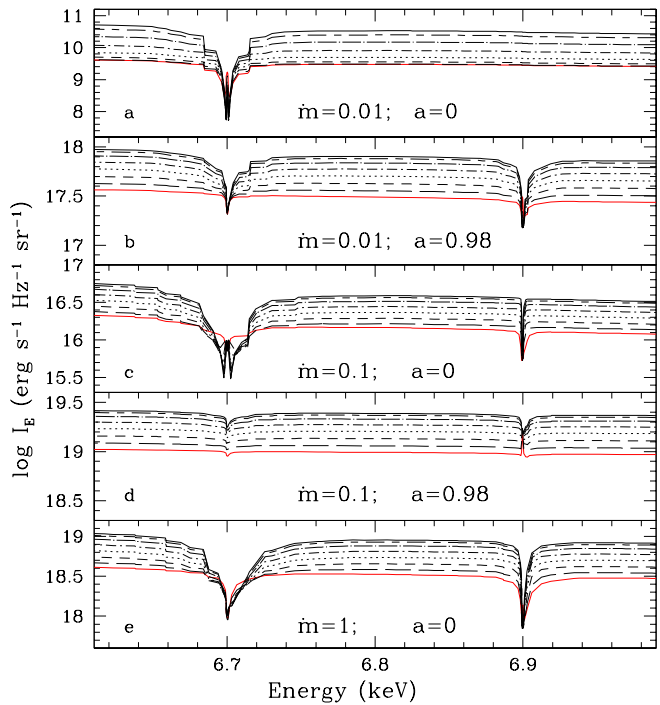}
\caption{Profiles of resonance lines from helium-like and hydrogen like-iron, integrated
over disk surface for all non-illuminated disk models. 
No kinematic either general relativity effects were taken into 
account.}
\label{fig:totli}
\end{figure}

\begin{figure}
\epsfxsize=8.8cm \epsfbox[80 150 580 700]{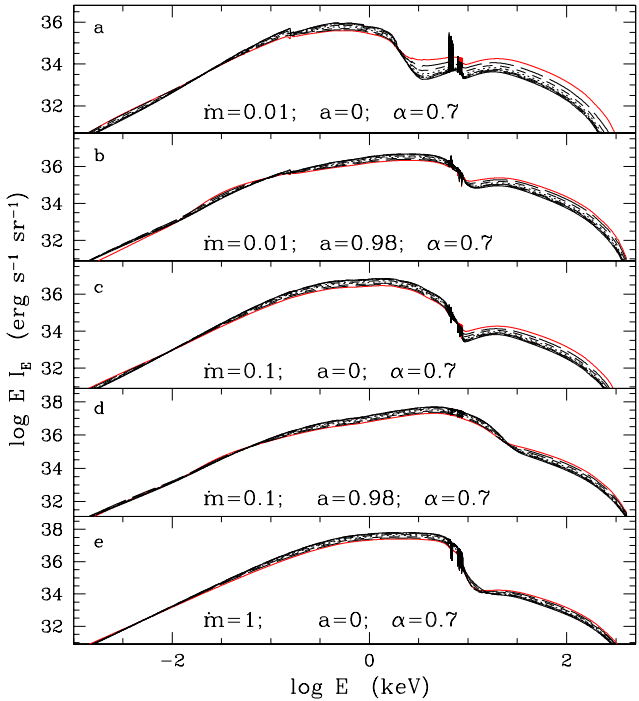 }
\caption{Broad-band intensity spectra integrated over disk surface 
for all irradiated disk models for eight viewing angles. The shape of continuum is 
extended toward hard X-rays due to both: thermal emission of hot atmospheric zone and
the radiation from the external source reflected by Compton scattering.
It is evident that continuum intensity spectra exhibit limb-brightening in 
X-ray domain.  }
\label{fig:totalosw}
\end{figure}

\begin{figure}
\epsfxsize=8.8cm \epsfbox[80 150 580 700]{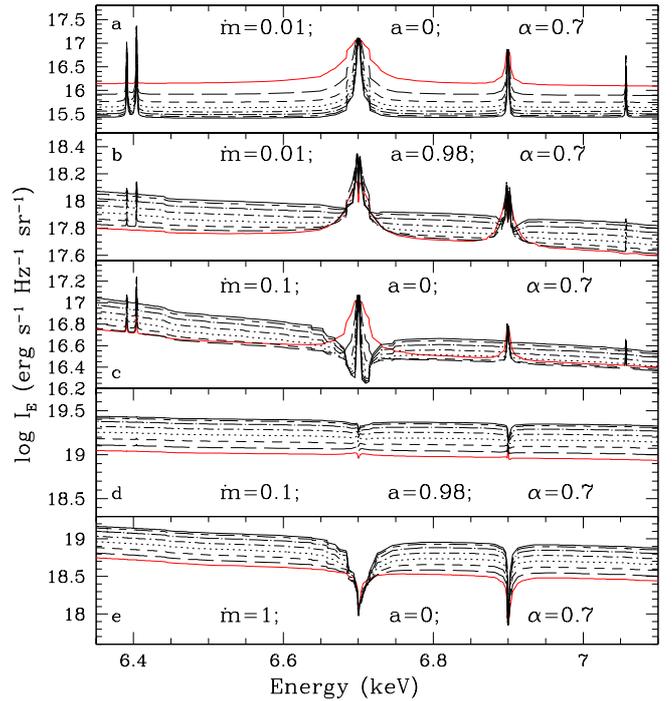}
\caption{Iron line complex in the energy range 6.3-7.2 keV, integrated
over disk surface for all illuminated disk models. In three upper panels, 
Lyman alfa lines of helium-like 
iron (6.7 keV) and of hydrogen-like iron (6.9 keV) are in emission, which is 
caused by the layer of temperature invertion in the uppermost atmosphere. 
Only for the hottest disk models (two bottom panels) both lines are in 
absorption, since the layer of temperature inversion is thin. 
Fluorescent emission lines, K$_{\alpha}$ at 6.4 keV and K$_{\beta}$ at 7.05 keV, 
can be seen only in colder disk models.}
\label{fig:totliosw}
\end{figure}

At the last step of computations we integrate 
spectra emitted from individual rings over disk surface 
according to the usual formula:
\begin{equation}
I_{E}^{tot}(\mu)=2 \pi \sum_{n=1}^{n=N}  I_E^{(n)} (\mu) \, R_n \, \Delta R_n \, ,
\end{equation} 
where $n$ is a ring number, $\mu$ is the cosine of viewing angle, and $\Delta R_n$ 
is the width of the $n$-th ring. 
  
Fig.~\ref{fig:total} presents continuum spectra integrated over disk surface for 
all non irradiated models. Left hand part of each panel displays featureless 
UV, optical and near infrared spectrum of disk whereas right hand part 
reaches hard X-rays and iron line region. 
All spectra show limb-darkening over the whole range of energies.   
This is clearly seen in Fig.~\ref{fig:totli}: intensity edge on (solid red line) 
is much lower than face on intensity (solid black line). 

Resonance iron lines appear always in absorption for non irradiated models
(Fig.~\ref{fig:totli}). They are very deep, i.e central intensity 
in the line can drop almost three orders of magnitude (panel a) as compared to the 
continuum level. Our code takes into account pressure 
broadening of all resonance lines. 
and particulary Ly${\alpha}$ 
line of He-like iron can be very wide, as seen in panels c and e. 
Such prominent lines certainly can be detected in present X-ray observations
\citep{kubota2007}.
Accretion disks  with highest effective temperatrue do not show
strong iron lines in absorption, 
and for the hotest model iron is fully ionized, as seen in panel d. 

Continuum spectra of the same disk models with external irradiation are 
displayed in Fig.~\ref{fig:totalosw}.
All spectra exhibit Compton reflection bump in hard X-rays. 
The relative position of Compton reflection in comparison with UV/X-ray disk bump 
depends on accretion rate and black hole spin. 
External irradiation causes shift to limb-brightening in the whole region of Compton 
reflection bump. 
For low accretion rate and non rotating black hole limb-brightening appears also 
in visible and near infrared spectral region 
(panels a and c in Fig.~\ref{fig:totalosw}).
 
Iron line complex for illuminated accretion disk models is presented in  
Fig.~\ref{fig:totliosw}. Structure of the iron line complex is
rather complicated and its properties, i.e. line absorption/emission and  
limb darkening/brightening, do not depend in the straightforward way 
on the accretion rate and the black hole spin. 

Lines from He-like and H-like iron appear in emission only for low accretion rates. 
For the accretion rate $\dot m=1$ those lines are in absorption. 
Higher level of irradiation may change this into emission, but this depends on 
the shape of the illuminating radiation. Increasing of soft X-ray photon number  
leeds to the efficient Compton down-scattering, which effectively 
cools down already hot atmosphere in galactic black hole binaries
(see Sec.~\ref{sec:temp}).   

Resonance lines in emission presented in panel b are very wide 
and do not show any limb brightening. We suppose that
this wide profile is caused by Compton scattering in hot atmosphere. 
As expected, model presented in panel d is featureless.

The important result of our paper is that in models with lower effective 
temperature, cold fluorescent K$_{\alpha}$ and K$_{\beta}$ lines are emitted. 
The strenght of these lines is large enough to contribute to the final 
iron line complex. In our code we assume that the cold fluorescent 
K$_\alpha$ line has fixed energy 6.4 keV. 
But this energy depends on detailed 
ionization structure of iron atoms, and we shall discuss this problem in a forthcoming paper.  

\section{Summary}
\label{sec:summary}

In this paper we present model atmospheres and intensity spectra
for a set of accretion slim disks around black holes in galactic black hole binaries.
Model atmospheres are computed for various accretion rates and 
spin of the black hole. 
Our computations include effects of Compton scattering and external 
irradiation by an X-ray lamp. Spectra are obtained for a large
number of individual rings, depending on the particular model, the 
average number is 14. 
Outgoing intensity spectra were computed for wide energy 
range from infrared to hard X-rays, up to 400 keV. 
Note, that results presented in our previous paper (RM08)
were calculated for SS disk global models, whereas the present 
paper deals with slim disks with advection. 

We demonstrate that the external irradiation developes huge temperature 
rise in the outermost layers of disk atmospheres. The effect is most 
prominent in cases when the atmosphere has lower effective temperature
(in the outer and the innermost rings of our slim disk models). 
In those colder rings, strong 
K$_{\alpha}$ and K$_{\beta}$ fluorescent lines of iron are emitted. 
Results of the paper demonstrate also that the external irradiation causes
a change of limb-darkening of non irradiated disk to limb-brightening 
particularly in hard X-ray region. 
This implies that accretion disks seen edge on are brighter in X-rays than 
disks seen face on. 

Our models clearly show that for high accretion rates 
it is rather difficult to obtain strong hot 
iron resonance lines in emission. This is due to the fact, that 
the disk atmospheres in galactic black hole binaries are generaly very hot. 
In such a situation iron can be fully ionized and produces no line features. 
For low accretion rates, resonance iron lines  appear in emission, when 
the heating by comptonization of external photons causes huge temperature rise.  

Our results differ from previously computed spectra 
from constant density slabs of 10 Thomson optical depth
\citep{ross2007,kallman2010}. 
This papers did not take into account  
the possibility of the resonance iron lines formation 
in hot atmospheres of the accretion disk around a stellar black hole.
In both cases iron lines appear only in emission.  

Results of this paper show that both resonance iron X-ray lines  
can be form in absorption for hot accretion disk atmopsheres in GBHc.
This is another possibility to explain observations 
of those absorption lines in some blak hole transients 
\citep[][and references therein]{kubota2007}.

Moreover, iron line profiles computed at source must be always pressure broadened, 
and their profiles deviate from the gaussian shape. 
Most of emission and absorption line profiles
are not gaussian (Fig.~\ref{fig:totli} and \ref{fig:totliosw}).
For the same accretion rate, iron line profiles differ for 
two values of spin considered in our paper (panel a, b, and c, c in 
Fig.~\ref{fig:totliosw}. 
This should be taken into account when fitting of black hole spin 
to observed the iron lines.

%
\begin{acknowledgements}
This research was supported by the Polish Ministry of Science 
and Higher Education grant No. N N203 511638.
\end{acknowledgements}

\bibliographystyle{aa}
\bibliography{refs}

\end{document}